\documentclass[aps,prl,superscriptaddress,showpacs,twocolumn]{revtex4}
\usepackage{graphicx}
\usepackage{color}
\usepackage{amssymb}

\begin{document}
\title{Low temperature magnetic transition in RuSr$_2$EuCeCu$_2$O$_{10}$ ruthenocuprate}

\author{I. \v{Z}ivkovi\'{c}}

\affiliation{Institute of Physics, Zagreb, Croatia}

\author{D. Paji\'{c}}

\affiliation{Physics Department, Faculty of Science, University of Zagreb, Zagreb, Croatia}

\author{K. Zadro}

\affiliation{Physics Department, Faculty of Science, University of Zagreb, Zagreb, Croatia}

\date{\today}
\begin{abstract}
A new magnetic transition in the ruthenocuprate parent compound RuSr$_{2}$EuCeCu$_{2}$O$_{10}$ has been observed below 10 K. It shows up only as a kink in the imaginary part of the ac susceptibility and exhibits a pronounced frequency dependence. At the same time, the real part of the ac susceptibility and the dc magnetization study show very little change in the same temperature window suggesting only a minor fraction of the material to be involved in the transition. Frequency dependence shows excellent agreement with the predictions of the Arrhenius law known to describe well the dynamics of the superparamagnetic particles. The same type of the investigation on the RuSr$_2$Eu$_{1.1}$Ce$_{0.9}$Cu$_2$O$_{10}$ composition showed no evidence of the similar transition, which points to a possible intrinsic behavior.
\end{abstract}

\maketitle

\section{Introduction}
Ruthenocuprate system RuSr$_{2}$\emph{RE}$_{2-x}$Ce$_{x}$Cu$_{2}$O$_{10}$ (Ru1222) (\emph{RE} = rare-earth element, $x = 0.4$ -- $1.0$) has been shown to exhibit peculiar magnetic behavior, the origin of which is still elusive. From the first reports~\cite{Bauernfeind1996} it has attracted much attention due to the possible coexistence of the ferromagnetic and superconducting order parameters. It is assumed that the magnetism originates from the RuO$_2$ planes, while the superconductivity is confined to the CuO$_2$ planes as in other high-T$_C$ materials. Later, the sister compound RuSr$_{2}$\emph{RE}Cu$_{2}$O$_{8}$ (Ru1212) has also revealed this possibility~\cite{Bernhard1999} and at the present there is a good amount of results from various techniques, including neutron diffraction~\cite{Lynn2000,McLaughlin2005}, muon spin rotation~\cite{Bernhard1999,Shengelaya2004}, dc and ac susceptibility~\cite{Williams2000,Felner2002,Zivkovic2002a,Zivkovic2002b} and M\"ossbauer spectroscopy~\cite{Felner2004}. However, this has not produced any definitive conclusion about the state of the magnetic order in ruthenocuprates. The main difficulty lies in the fact that the results from the magnetization~\cite{Felner2002}, M\"ossbauer~\cite{Felner2004}, ESR~\cite{Yoshida2003} and $\mu $SR~\cite{Shengelaya2004} suggest the existence of the ferromagnetic component with saturation moment around $1 \mu _B$ while the neutron diffraction has shown antiferromagnetic long-range order~\cite{Lynn2000,McLaughlin2005} with a possible ferromagnetic component $< 0.1 \mu _B$. Xue and coworkers have suggested the phase-separation scenario for both compositions~\cite{Xue2003,Xue2003a} with the ferromagnetic islands inside the antiferromagnetic matrix which would somewhat reconcile two opposing hypotheses. Evidence that at high temperatures only about 15\% of the volume is ordered in Ru1222 system~\cite{Shengelaya2004,Felner2004} supports this scenario but the microscopic picture is still lacking.

Recently, a new compound in the ruthenocuprate family with the formula RuSr$_{2}$RECe$_{2}$Cu$_{2}$O$_{12}$ (Ru1232) has been synthesized~\cite{Awana2005}. It shows similar magnetic behavior as well as superconductivity. No detailed investigation has been published so far on this compound.

In this Report we focus our attention on the Ru1222 system, particularly on the parent composition RuSr$_{2}$\emph{RE}$_{2-x}$Ce$_{x}$Cu$_{2}$O$_{10}$ with $x = 1.0$. It is not superconducting because stoichiometricaly the copper ions are in the 2+ state (no doped holes) and Ru ions are in the 5+ state. On the other hand it should be less permeable to the oxygen inhomogeneities which are unavoidable for the $x < 1.0$~\cite{McLaughlin2003,Awana2003}. In previous report~\cite{Felner2002} the existence of a magnetic transition around $T_M^F =180$ K has been shown, below which the zero field cooled (ZFC) and field cooled (FC) magnetization curves start to branch, indicating a development of an ordered phase. Below this temperature $\mu $SR~\cite{Shengelaya2004} and M\"ossbauer spectroscopy~\cite{Felner2004} revealed that only 15\% of the sample volume is ordered. The main magnetic transition occurs at $T_M = 120$ K~\cite{Felner2002,Zivkovic2002b} where ac and dc susceptibilities start to rise sharply indicating some kind of a ferromagnetic transition, although frequency dependence of the peak has been observed and the spin-glass order suggested~\cite{Cardoso2003}. Below $T_M$ the time relaxation of ac susceptibility and the inverted butterfly hysteresis has been observed~\cite{Zivkovic2002b,Zivkovic2006}. From $\mu $SR~\cite{Shengelaya2004} and M\"ossbauer spectroscopy~\cite{Felner2004} a 100\% of the sample volume is in the ordered state. For the compositions with $0.4 < x < 0.8$ there is a superconducting transition around 30 K.

Alongside this main features, an anomaly around 130 K has been observed but it has been argued that it is not a bulk transition~\cite{Shengelaya2004}. Also, between the main peak at $T_M$ and the superconducting transition Cardoso and coworkers noticed thermal hysteresis with the peak in the imaginary part of the ac susceptibility for the $x = 0.5$ composition which shifts with the frequency~\cite{Cardoso2005}. They ascribed it to the rearrangement of the spins within the spin-glass state.

We have performed detailed investigation of the real and imaginary part of the ac susceptibility in the Ru1222 system at low temperatures. Here we present, to the best of our knowledge, for the first time the low temperature transition in Ru1222 system.

\section{Experimental details}

We have investigated two different batches of the ruthenocuprate RuSr$_{2}$\emph{RE}$_{2-x}$Ce$_{x}$Cu$_{2}$O$_{10}$ with $x = 1.0$ composition which were prepared at different times following the solid-state reaction procedure. X-ray diffraction measurements indicated that the impuritiy levels are very small ($\sim 2\%$)~\cite{Zivkovic2002b}. We have denoted the samples with S1 and S2. The signal from the samples didn't change through several years, indicating good stability of the composition. Ac susceptibility measurements were taken by the use of a commercial CryoBIND system. All the measurements were conducted in the heating regime with a heating rate below 2 K/min. Magnetization measurements were performed using a commercial Quantum Design MPMS5 SQUID magnetometer. Samples were cooled down from 250 K to 2 K in zero field, then the measuring field would be applied and the ZFC magnetization curve measured up to 250 K. After cooling the sample in the presence of the field, the FC curve was measured while increasing the temperature.

\section{Experimental results}

\begin{figure}
\begin{center}
\includegraphics{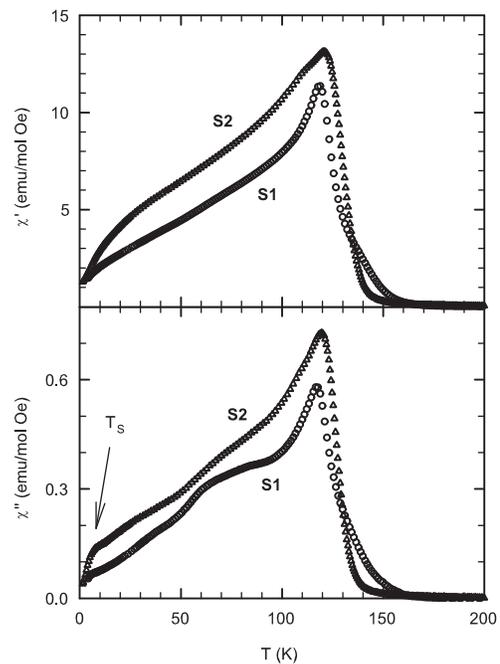}
\caption{Temperature dependence of the real (upper panel) and the imaginary part (lower panel) of the ac susceptibility for two samples of the Ru1222EuCe composition. Frequency and the amplitude of the measuring field were 431 Hz and 0.1 Oe, respectively.} \label{temp_dependence}
\end{center}
\end{figure}

In Fig.~\ref{temp_dependence} we present temperature dependence of the real and imaginary part of the ac susceptibility for two samples of the Ru1222 system with the $x = 1.0$ composition.
Both samples show similar behavior, including the main peak at $T_M = 120$ K, a shoulder around 60 - 70 K~\cite{Cardoso2003,Cardoso2005} (visible in the imaginary part) and a drop in the real part below 30 K accompanied with a kink in the imaginary part around 10 K ($T_S$). For the S1 sample there is a visible signature of the anomaly around 130 K that is not present for the S2 sample. Taking into account that below 130 K both curves appear almost identical, we can conclude that the anomaly does not influence basic magnetic ordering in the Ru1222 system. This conclusion is in agreement with observations from $\mu $SR that the anomaly does not show a bulk character~\cite{Shengelaya2004}.

Worth mentioning, S2 sample shows somewhat larger signal in both real and imaginary part of the ac susceptibility below $T_M$ but eventually at the lowest temperatures the signals become equal in magnitude, suggesting a similar ground state.

\begin{figure}
\begin{center}
\includegraphics{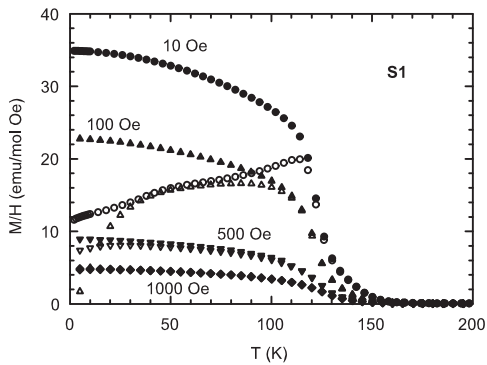}
\caption{Magnetization vs. temperature for the S1 sample. Open symbols represent the ZFC curves while the full symbols represent FC curves. Respective FC curves are labeled with the measuring field.} \label{squid}
\end{center}
\end{figure}

DC magnetic measurements of the S1 sample are shown in Fig.~\ref{squid}. For small fields around $T_M$ there is a strong increase in the signal below which a cusp in the ZFC curve develops and the branching between FC and ZFC curves starts to grow. This has been attributed to the emergence of the weak ferromagnetic state~\cite{Felner2002}. The FC--ZFC divergence is present at temperatures as high as 200 K, around the temperature where $\mu $SR and M\"{o}ssbauer spectroscopy revealed the ordering of the minority fraction~\cite{Shengelaya2004,Felner2004}. Around 50 K there is a kink in the ZFC curves also observed by Cardoso and coworkers~\cite{Cardoso2003}. ZFC curve measured in 10 Oe does not show a drop indicating that the sample was cooled not in zero but in a field with finite value. We ascribe this effect to the remanent field in the superconducting magnet. As the field increases, the cusp in the ZFC curve broadens and moves to lower temperatures. For $H = 1000$ Oe there is no difference between the FC and ZFC curves.

\begin{figure}[b]
\begin{center}
\includegraphics{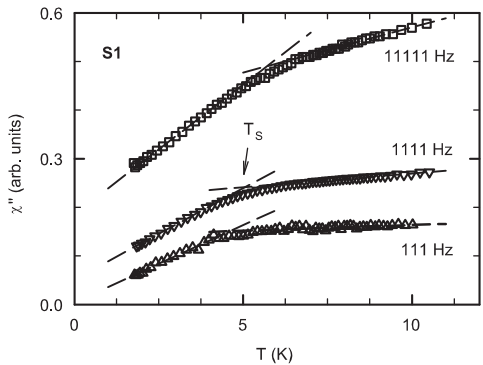}
\caption{Frequency dependence of the imaginary part of the ac susceptibility in the low temperature region for the S1 sample. The amplitude of the ac magnetic field was 0.1 Oe. To improve the clarity, only a small portion of the experimental points is shown.} \label{low_S1}
\end{center}
\end{figure}

\begin{figure}[b]
\begin{center}
\includegraphics{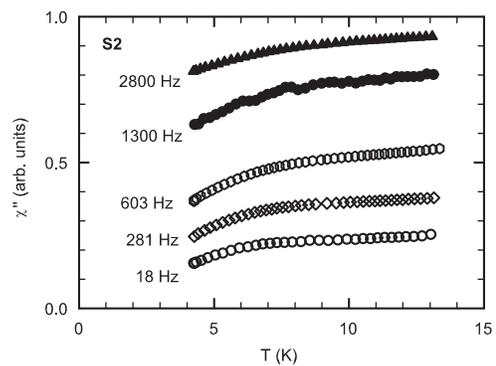}
\caption{Frequency dependence of the imaginary part of the ac susceptibility in the low temperature region for the S2 sample. The amplitude of the ac magnetic field was 1 Oe. To improve the clarity, only a small portion of the experimental points is shown and few frequency curves have been omitted (28 Hz, 61 Hz, 131 Hz).} \label{low_S2}
\end{center}
\end{figure}

Now we proceed to the main result of this Report: the low temperature transition in the Ru1222EuCe system. We have investigated the frequency dependence of the ac susceptibility of the Ru1222EuCe system and in the Figs.~\ref{low_S1} and~\ref{low_S2} we show the relevant temperature region around 10 K of the imaginary part for the S1 and S2 samples, respectively.
As the frequency of the driving field is increased, there is an overall shift of the signal to higher values, indicating greater energy dissipation in the sample. For both samples there is a kink in the curves which shifts to higher temperatures as the frequency is increased. Due to the roundness of the kink, a straightforward determination of the characteristic temperature is not possible. Therefore, for each frequency, we have fitted linear parts of the curve below and above the kink and then used the crossing of the two lines to define $T_S$, as shown in Fig.~\ref{low_S1}. There is a small difference in obtained $T_S$ for two samples. In the investigated frequency range, S1 exhibits a kink between 4 K and 6 K, while the kink for S2 spans from 6 K to 8 K. In the same temperature window the real part of the ac susceptibility shows a minor change in the slope of the curve and varies only slightly with frequency (not shown).

\begin{figure}
\begin{center}
\includegraphics{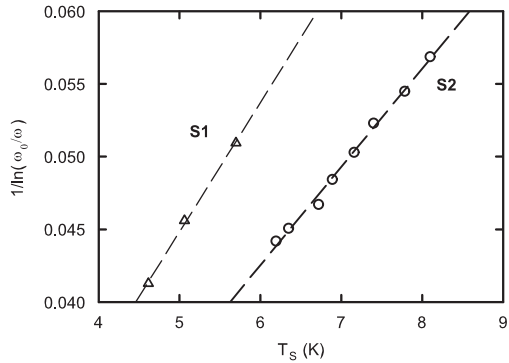}
\caption{The frequency dependence of the characteristic frequency $T_S$ for the S1 and S2 samples. The dashed lines are best fits using the Arrhenius law (see text).} \label{arrhenius}
\end{center}
\end{figure}

After extracting the characteristic temperature for each frequency, we can search for an appropriate description of the frequency dependence $T_S (\omega )$. A quantitative measure of the frequency shift can be obtained from the ratio $\Delta T_S/(T_S \Delta \log \omega)$~\cite{Mydosh1993}. Cardoso and coworkers used this ratio as a signature of the presence of the spin-glass order below the main magnetic transition at $T_M$. In our case, this ratio is in the range of 0.1 which is somewhat lower than the characteristic superparamagnetic value but substantially larger than the spin-glass values which go as low as 0.005 (\emph{Cu}Mn) but can be as high as 0.06 (Eu$_{1-x}$Sr$_x$S)~\cite{Mydosh1993}. In Fig.~\ref{arrhenius} we show the frequency dependence of the characteristic temperature $T_S$ through the Arrhenius law

$$\omega = \omega_0 e^{-\frac{U}{k_B T_S}},$$

known to describe well the superparamagnetic systems~\cite{Bean1959}. Here, $\omega = 2 \pi \nu$ is the frequency of the magnetic field, $\omega _0$ is the attempt frequency and $U$ is the potential barrier between the two easy orientations of the superparamagnetic particle.
Extracted parameters of the Arrhenius law are $\nu_0 = 3.7 \cdot 10^{12}$ Hz, $U = 112$ K and $\nu_0 = 1.2 \cdot 10^{11}$ Hz, $U = 142$ K for the S1 and S2 samples, respectively. The dashed lines in Fig.~\ref{arrhenius} represent the best fits using the above values.

\section{Discussion}

The main question to be answered here is to what kind of a transition does the above described event belongs to. As shown above it displays a pattern familiar for blocking of the superparamagnetic particles. Here, spins on the magnetic ions are connected through some kind of an internal interaction into a cluster whose total spin fluctuates at high temperatures but is blocked in the easy axis direction at lower temperatures. Intrinsic to this scenario is a phase separation between the superparamagnetic particles and the matrix in which these particles are dispersed. In ruthenocuprates a phase separation has been proposed to account for a variety of conflicting measurements~\cite{Shengelaya2004,Felner2004,Xue2003a}. Specifically, $\mu $SR~\cite{Shengelaya2004} and M\"{o}ssbauer spectroscopy~\cite{Felner2004} showed that only 15\% of the sample gets ordered around 180 K, while the rest of the volume orders at the main magnetic transition temperature $T_M$. In that context, the presence of another species of superparamagnetic particles in the system should not be neglected. Relatively small change in the real part of the ac susceptibility below the blocking temperature suggests only a minor volume fraction and explains why it went unnoticed so far.

There is also a question about the origin of the superparamagnetic particles, i.e. are they intrinsic to the material or impurity related. Recently, Felner and coworkers investigated the origins of the high temperature phase separation and suggested either high concentration regions of Ru$^{4+}$ ions or minor SrRu$_{1-y}$Cu$_y$O$_3$ magnetic phase to be responsible for the formation of the ordered fraction~\cite{Felner2005}. Additionally, they showed that the parent composition with $x = 1.0$ is intrinsically different with respect to the samples with less Ce. For $x < 1.0$ there is a depletion of oxygen which compensates the charge disbalance caused by an increase in the \emph{RE}/Ce ratio. The lack of oxygen ions can be inhomogeneously distributed throughout the material, inducing the phase separation. Interestingly, the investigation of the $x = 0.9$ composition (the only non-superconducting composition besides the $x = 1.0$; not shown), prepared at the same time as the S1 sample, shows no evidence for a superparamagnetic blocking similar to the one described in this Report for the $x = 1.0$ composition. Also, the fact that we have observed the same feature in two different samples of the same $x = 1.0$ composition prepared at different time suggests that this is not related to a specific batch. This leads us to conclude that the origin of the superparamagnetic particles could be intrinsically connected to the parent composition with $x = 1.0$.

\section{Conclusion}

In the variety of magnetic features known to exist in the ruthenocuprate family we have observed another one. Below 10 K there is a kink in the imaginary part of the ac susceptibility for the Ru1222EuCe samples which shows frequency dependence characteristic of the blocking of the superparamagnetic particles. The real part of the ac susceptibility and dc measurements show minor change in the same temperature window. Only the parent compound with $x = 1.0$ composition reveals this transition while in the $x = 0.9$ composition it is absent. This could be a good indication of the intrinsic behavior in the material beacuse it has been shown that compositions with $x < 1.0$ are affected by the change in the oxygen content and the Ru valence differs from the 5+ value~\cite{McLaughlin2003,Awana2003}. Since only two samples have been investigated, an impurity related conclusion should not be discarded as yet.

\section{Acknowledgments}

We thank prof. I. Felner for providing us with samples. I. \v{Z}. acknowledges discussions with M. Prester and D. Drobac. This work was supported by the resources of the SNSF-SCOPES project.



\end{document}